\documentclass{sig-alternate}

\begin{document}

\newtheorem{definition}{Definition}

\newcommand{\CDIRECT}{E[d_A(t_A)]}
\newcommand{\CFUT}{E[o_A(t_A)]}
\newcommand{\REV}{E[r_A(t_A)]}
\newcommand{\LENGTH}{t}
\newcommand{\CREMV}{v_A(i)}
\newcommand{\CSETUP}{u_A}
\newcommand{\DIFF}{\theta_A(i,t_A)}
\newcommand{\DIFFA}{\theta_A(a,t_A)}
\newcommand{\CEXEC}{e_A(t_A)}
\newcommand{\CEFF}{f_F(i)}
\newcommand{\CREMVF}{v_F(i)}
\newcommand{\CREMVX}{v_x(i)}
\newcommand{\ECONOMICS}{q_A}

%%% Mittelwert-Notation mit Quer
\newcommand{\MREMV}{v_A(\TYPE)}
\newcommand{\MREMVX}{v_x(\TYPE)}
\newcommand{\MEXEC}{e_A(t_A)}
\newcommand{\MEXECX}{e_x(t_x)}
\newcommand{\MEFF}{f_F(\TYPE)}
\newcommand{\MSETUP}{u_A}
\newcommand{\MSETUPX}{u_x}
\newcommand{\MREMVF}{v_F(\TYPE)}

%%% Defect classes and types
\newcommand{\TYPE}{\tau_i}
\newcommand{\TYPEJ}{\tau_j}
\newcommand{\PTYPE}{p_i(\TYPE)}
\newcommand{\CLASSES}{C}
\newcommand{\CLASS}{c}
\newcommand{\ECLASS}[1]{E_{p_i^c}(#1)}
\newcommand{\DDT}{de\-fect-de\-tec\-tion tech\-nique}
\newcommand{\DIFFT}{\theta_A(\TYPE,t_A)}

\newcommand{\PCLASS}{q_i(\CLASS)}

%%% E-Notation derzeit nicht genutzt
\newcommand{\EREMV}{E_\alpha(i,\CREMV)}
\newcommand{\EEFF}{E_\sigma(i,\CEFF)}
\newcommand{\EREMVF}{E_\alpha(\CLASS,\CREMVF)}
\newcommand{\EREMVX}{E_\alpha(i,\CREMVX)}

%
% --- Author Metadata here ---
\conferenceinfo{SOQUA'06,}{November 6, 2006, Portland, OR, USA.}
\CopyrightYear{2006} % Allows default copyright year (2000) to be over-ridden - IF NEED BE.
\crdata{1-59593-584-3/06/0011}  % Allows default copyright data (0-89791-88-6/97/05) to be over-ridden - IF NEED BE.
% --- End of Author Metadata ---

\title{Integrating a Model of Analytical Quality Assurance into the V-Modell XT}
%
% You need the command \numberofauthors to handle the "boxing"
% and alignment of the authors under the title, and to add
% a section for authors number 4 through n.
%
% Up to the first three authors are aligned under the title;
% use the \alignauthor commands below to handle those names
% and affiliations. Add names, affiliations, addresses for
% additional authors as the argument to \additionalauthors;
% these will be set for you without further effort on your
% part as the last section in the body of your article BEFORE
% References or any Appendices.

\numberofauthors{1}
%
% You can go ahead and credit authors number 4+ here;
% their names will appear in a section called
% "Additional Authors" just before the Appendices
% (if there are any) or Bibliography (if there
% aren't)

% Put no more than the first THREE authors in the \author command
\author{
%% e-mail address with \email.
\alignauthor Stefan Wagner and Michael Meisinger\\
       \affaddr{Institut f\"ur Informatik}\\
       \affaddr{Technische Universit\"at M\"unchen}\\
       \affaddr{Boltzmannstr.\ 3, D-85748 Garching b.\ M\"unchen, Germany}\\
       \email{\{wagnerst,meisinge\}@in.tum.de}
}

\maketitle

\begin{abstract}
Economic models of quality assurance can be an important tool 
for decision-makers
in software development projects. They enable to base quality
assurance planning on economical factors of the product and the
used defect-detection techniques. A variety of such models
has been proposed but many are too abstract to be used in practice.
Furthermore, even the more concrete models lack an integration
with existing software development process models to increase their applicability.
This paper describes an integration of a thorough stochastic model
of the economics of analytical quality assurance with the systems development process
model V-Modell XT.
The integration is done in a modular way by providing a new process module
 -- a concept directly available in the V-Modell XT for extension 
purposes -- related to analytical quality assurance.
In particular, we describe the work products, roles, and activities defined in our new
process module and their effects on existing V-Modell XT elements. 
\end{abstract}

\category{D.2.9}{Software Engineering}{Management}
%A category including the fourth, optional field follows...
\category{D.2.8}{Soft\-ware Engineering}{Metrics}
\category{D.2.5}{Software Engineering}{Testing and Debugging}

\terms{Economics, Verification, Reliability}

\keywords{Software quality economics, quality costs, quality model, quality
          assurance, process model, V-Modell}

\section{Introduction}
\label{sec:intro}

Software quality costs and economics have been subject to research for
decades now. Consequently, there is a variety of corresponding models
on all levels of abstraction as a result of this research. The
development and improvement of these models is important, especially
for the decision makers in real software projects \cite{rai98}.
This becomes obvious when considering that there are many estimates
that assign 30--50\% of the development costs to quality assurance
\cite{myers79,jones91}.  A newer study of the National 
Institute of Standards
and Technology of the United States \cite{nist02} found that even 80\% of the
development costs are caused by the
detection and removal of defects. Hence, models are needed
to control and minimise these costs. Yet, for this to be feasible we need
to incorporate them into existing development processes. Thereby, we make them
operational and accessible for decision-makers.

\subsection{Problem}
Most models are not directly applicable in a real
development process. They often only classify the relevant costs but
do not show how to use this classification. Even operational models
often neglect the fact that they need to be used in the context of
a specific process model. The usage of such models is mainly in
an ad-hoc manner and they are not systematically included in
process models.

\subsection{Contribution}
The contribution lies in the seamless integration of a model for
analytical software quality assurance into the existing 
process model V-Modell~XT. We show how our QA model is operationally
used and which roles, products, and activities are involved in
using the model in practice. This allows an easy adoption of the model
for a project that follows the V-Modell~XT. Although similar ad-hoc
usages of such models are practice in some companies, we are not aware
of an earlier systematic integration.

\subsection{Outline}
First, we introduce quality economics in general and in terms of
the analytical model in Sec.~\ref{sec:model}. In Sec.~\ref{sec:vmodell}
we describe the basics of the considered process model and its
underlying meta-model. Sec.~\ref{sec:integration} then shows the
integration of the model with the V-Modell XT. We finish with related
work in Sec.~\ref{sec:related} and final conclusions in 
Sec.~\ref{sec:conclusions}.

\section{Quality Economics}
\label{sec:model}

We first describe the cost types and other factors that are
important in the context of the economics of analytical quality
assurance. Then we give a short overview of the analytical model
from \cite{wagner:issta06,wagner:diss06} that is to be integrated
in the process model V-Modell~XT later.

\subsection{Cost Types and Factors}

\begin{figure}[h]
\begin{center}
  \includegraphics[width=.45\textwidth]{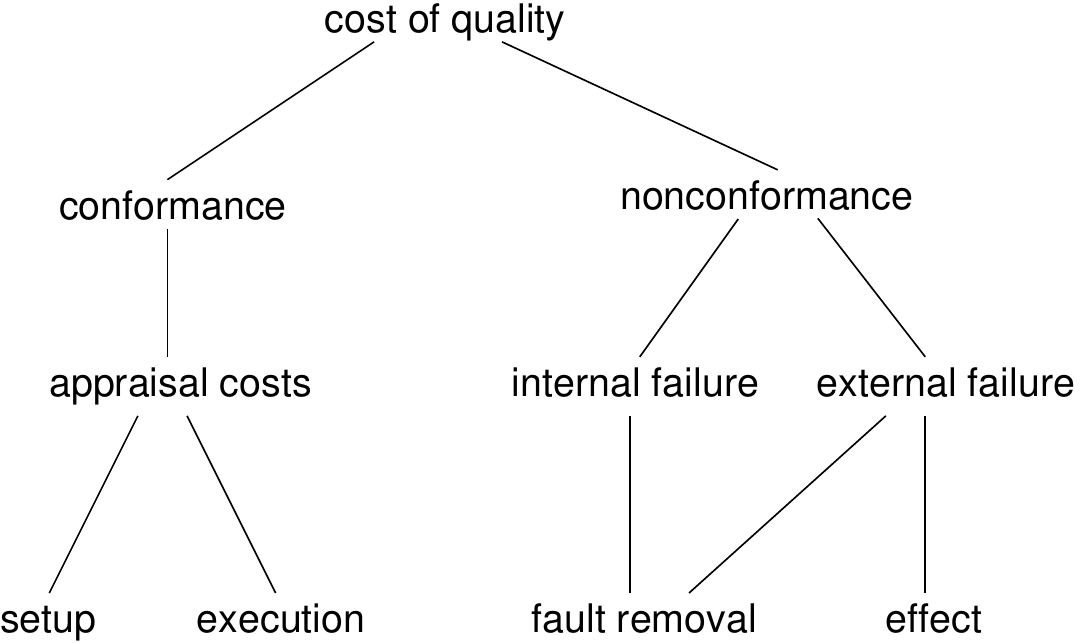}
  \caption{The refined cost types}
  \label{fig:costs_revisited}
\end{center}
\end{figure}

We reduce the classical PAF (Prevention, Appraisal, Failure) model 
of quality costs 
to an AF (Appraisal, Failure) 
model. 
We ignore \emph{prevention costs} that contain the costs of preventing
defects by constructive QA because constructive QA has significantly 
different characteristics.
\emph{Appraisal costs} contain all costs for checking artefacts
to detect defects, e.g., test specification and execution. The debugging
is then part of the failure costs. When the failure occurs in-house it
incurs \emph{internal failure costs}. Failures during operation at
the customer cause \emph{external failure costs}. We refine these 
parts
so that we can identify the relevant cost factors.
The complete refined model is shown in Fig.~\ref{fig:costs_revisited}.
The appraisal costs
are detailed to \emph{setup} and \emph{execution} costs. The former
constituting all
initial costs for buying test tools, configuring the test environment,
and so on. The latter includes costs that are connected to actual
test executions or review meetings, mainly personnel costs.

On the nonconformance side, we have \emph{fault removal} costs that can
be attributed to the internal failure costs as well as the external
failure costs. This is because if we found a fault and wanted to remove it,
it would always result in costs no matter whether caused by an internal
or external failure. Actually, there does not have to be a failure at all.
Considering code inspections, faults are found and removed that have never
caused a failure during testing. 
Fault removal costs also contain the costs for necessary re-testing and
re-inspections.

External failures also cause \emph{effect} costs. Those are all further costs
associated with the failure apart from the removal costs. 
For example, compensation
costs could be part
of the effect costs, if the failure caused some kind of damage
at the customer site. We might also include other costs such as loss of 
sales because of bad
reputation in the effect costs.

Furthermore, there are also technical factors that are important for
the quality economics of analytical quality assurance. The two main
factors that we consider in the following model are (1) the difficulty
of defect-detection and (2) the failure probability of faults. We denote
the probability that a specific defect-detection technique does not detect
a defect of a specific type as its difficulty. This factor has been shown
to be influential \cite{wagner:issta06}. A smaller but still substantial
impact stems from the failure probability of faults. This is important
because many faults occur only with a very small probability during operation
\cite{boehm01}.

\subsection{Analytical Model}

We use the stochastic quality assurance model from
\cite{wagner:issta06,wagner:diss06} as the model to be integrated
in the V-Modell~XT. Actually, we only consider the \emph{practical}
model from this work because that is the one to be applied. 
It is derived from a \emph{theoretical} model that
incorporates more factors and more detail. However, the main factors
described above are still contained in this model.

The main idea of the model is to compute the expected values
of the costs and benefits of quality assurance. For this purpose,
they are structured in three components: direct costs, future
costs and revenues. The determination of the expected values is
based on average values calculated from literature and finished
projects. During the project, measured data can be used to refine
the results.

We define $\TYPE$ to be the defect type of fault $i$. It is determined
using the defect type distribution of older projects.
In this way we do not have to look at individual
faults but analyse and measure defect types for which the determination
of interesting  quantities is possible during quality assurance. We will not
further elaborate the concept of defect types but refer to
defect classification approaches from IBM \cite{kan02}
or HP \cite{grady92}. For the sake of a simple presentation, we first
give equations for a single defect-detection technique and generalise
that to a combination of techniques.

\subsubsection{Single Economics}

We start with the direct costs $d_A$ of a \DDT . They are all costs that occur
directly by using the technique.

\begin{equation}
  \label{eq:direct_practical}
  \CDIRECT = \MSETUP + \MEXEC
             + \sum_{i}{
             (1 - \DIFFT) \MREMV},
\end{equation}
where $\MSETUP$ is the average setup cost for technique $A$, $\MEXEC$ is
the average execution cost for $A$ with effort $\LENGTH$, and $\MREMV$
is the average removal cost for defect type $\TYPE$. 

The future costs $o_A$ are those costs that will occur when defects are
not detect by the technique.
\begin{equation}
  \label{eq:future_practical}
  \CFUT =
  \sum_i{\pi_{\tau_i} \DIFFT (\MREMVF + \MEFF)}.
\end{equation}

Finally, the revenues $r_A$ are the saved future costs, i.e., the costs that
will not incur because the technique finds them.
\begin{equation}
  \label{eq:saved_practical}
  \REV =
  \sum_i{\pi_{\tau_i} (1 - \DIFFT)(\MREMVF + \MEFF)},
\end{equation}
where $\MEFF$ is the average effect costs of a fault of type $\TYPE$.

\subsubsection{Combined Economics}

The extension to more than one technique needs to consider whether
the defects have been found by earlier used techniques.
The following is the equation for the expected value of the direct
costs:
\begin{equation}
\begin{split}
\label{eq:practical_direct_combined}
  E[d_X(t_X)] = \sum_{x \in X}{\biggl[} & \MSETUPX + \MEXECX + 
       \sum_i{\Bigl[
       (1 - \theta_x(\TYPE,t_x))} \\
       & \cdot \prod_{y < x}{\Bigl( \theta_y(\TYPE,t_y)}
       \Bigr) \MREMVX  \Bigr] \biggr],
\end{split}
\end{equation}
where $X$ is the ordered set of the used \DDT s.
Also the expected value of the combined future costs $o_X$ can
be formulated in the practical model using defect types.
\begin{equation}
\begin{split}
  E[o_X(t_X)] = \sum_i{\biggl[} & \pi_{\TYPE} 
                \prod_{x \in X}{\Bigl(\theta_x(\TYPE,t_x)
                \Bigr)} \\
                & \cdot \Bigl( \MREMVF + \MEFF \Bigr) \biggr]
\end{split}
\end{equation}
Finally, the expected value of the combined revenues $r_X$ are defined
accordingly.
\begin{equation}
\begin{split}
\label{eq:practical_revenues_combined}
  E[r_X(t_X)] = \sum_{x \in X}{\sum_i{\biggl[}} & \pi_{\TYPE}
       (1 - \theta_x(\TYPE,t_x))
        \prod_{y < x}{\Bigl( \theta_y(\TYPE,t_y)}
                             \Bigr) \\
        & \cdot \Bigl( \MREMVF + \MEFF \Bigr) \biggr]
\end{split}
\end{equation}

\subsubsection{Needed Quantities}
\label{sec:needed}

Using the practical model, we identify only seven different types of quantities
that are needed to use the model:

\begin{itemize}
\item Estimated number of faults: $I$
\item Distribution of defect types
\item Difficulty functions for each technique and type $\theta_x(\tau_i)$
\item Average removal costs for type $\TYPE$ with technique $x$: $\CREMVX$
\item Average removal costs for type $\TYPE$ in the field: $\CREMVF$
\item Average effect costs for type $\TYPE$ in the field: $\CEFF$
\item Failure probability of fault of type $\TYPE$: $\pi_{\tau_i}$
\end{itemize}

For an early application of the model, average values from a
literature review can be used as first estimates. We did an extensive
analysis of those values in \cite{wagner:isese06} and ranked them
using sensitivity analysis in \cite{wagner:issta06}.
For more specific estimations we can use more sophisticated methods:
The 
COQUALMO model \cite{chulani99} allows to determine an estimate
of the number of faults contained in the software. 
The defect
removal effort for different defect types can be predicted using
an association mining approach of Song et al.\ \cite{song06}.

\subsubsection{Optimisation}
\label{sec:optimisation}

Optimisation is the key to using the model operationally in a
project. It allows to calculate an optimal effort distribution
for the used \DDT s.
Only two of the three components of the model
are needed because the future costs and the revenues are dependent
on each other. There is a specific number of faults that have associated
costs when they occur in the field. These costs are divided in the two
parts that are associated with the revenues and the future costs, respectively.
The total always stays the same, only the size of the parts varies
depending on the defect-detection technique. Therefore, we use only the 
direct costs and the revenues
for optimisation and consider the future costs to be dependent on the
revenues.

Hence, the optimisation problem can be stated by: maximise $r_X - d_X$.
By using Eq.~\ref{eq:practical_direct_combined} and 
Eq.~\ref{eq:practical_revenues_combined} we get the following term to
be maximised.
\begin{equation}
\begin{split}
\sum_x{\biggl[ - \MSETUPX - \MEXECX + \sum_i{\sum_{\CLASS}{\Bigl( \PCLASS
(1 - \theta_x^{t_x}) }}} \cdot \\
\phantom{\sum{\biggl[}}\prod_{y < x}{(\theta_y(c)^{t_y})} \bigl(\pi_i \MREMVF + 
\pi_i \MEFF -
 \MREMVX \bigr)
\Bigr) \biggr]
\end{split}
\end{equation}

For the optimisation purposes, 
we usually also have some restrictions, for example a maximum effort
$t_{\textit{max}}$ with $\sum_x{t_x} \leq t_{\textit{max}}$, either fixed
length or none $t_A = \{0,100\}$, or some
fixed orderings of techniques, that have
to be taken into account. The latter is typically true for different forms
of testing as system tests are always later in the development than unit
tests. We assume there is sufficient tool-support available to solve this
optimisation problem

\section{V-Modell XT}
\label{sec:vmodell}

The V-Modell XT~\cite{v-modell2006,Rausch2006} is a recently released German software and system development standard. It covers all relevant management, engineering and supporting processes of software development, for instance project management, quality assurance, offer, bidding and contract management, and also technical disciplines such as requirements engineering, system design and integration, software development and more specific engineering activities.

The goals of the V-Modell XT are to provide a generic development process model, which is easy to understand and to use, flexibly adaptable to the needs of organisations and projects, and reproducibly leading to developed products of higher quality with less cost and resources spent.

\subsection{Concepts of the V-Modell XT}

In order to extend the V-Modell XT it is imperative to know its main concepts. The V-Modell XT is based on a rigorous meta-model, which defines all concepts and their relationships. The entire process model strictly follows this meta-model, which is a prerequisite for flexible extensibility. The main concepts of the V-Modell XT are:

\begin{itemize}
\item \textit{Work Products} are the main project results and artefacts 
      (documents, models, code, deliverable systems). They have a defined 
      structure and prescribed content, and can be structured further into specific subjects (sub-sections). Work Products have a responsible creator and will be quality checked. An evaluation specification defines the requirements for their quality.
	
	\item \textit{Product Dependencies} define the consistency relations between the contents of different work products. Adhering to and checking product dependencies makes sure that all work products in a project will be created and kept consistent to existing products. Product dependencies are an important means to assure overall product quality and to trace information across products, for instance from requirements to software architecture elements.

	\item \textit{Activities} define the actions that need to be performed in order to create the work products. Activities can be structured further into sub-activities. Activities provide support for the actual doing within a project. There is exactly one activity per work product. In case there are multiple instances or iterations of a certain work product, the respective activity is performed several times, accordingly. Each activity creates a work product; therefore it is directly followed by a product evaluation for QA, if required.

	\item \textit{Roles} describe profiles of responsibility for the people working in the project. Roles will be impersonated by specific people in the project.
	
	\item \textit{Process Modules} group together Work Products, Activities and Roles, as well as other V-Modell~XT elements into self-contained units covering certain project processes, such as project management, requirements management, systems integration, software development, etc. Process modules have a hierarchy of dependence. They can be understood, applied and modified independently if adhering to these dependencies. Process modules are the main units of tailoring and extension of the V-Modell~XT. Important for the V-Modell~XT extension mechanism is that process modules can define work products, subjects, activities, sub-activities etc.\ that modularly and seamlessly extend existing processes of the core process modules.
	
	\item \textit{Tailoring} is the process of adapting the V-Modell~XT to a specific project or organisation. Tailoring consists of selecting the appropriate process modules out of the repository of available ones. After the tailoring, a consistent and adapted software development process exists. It is indistinguishable to the user from which process modules which specific elements -- work products, their subjects, extensions etc -- came from.
	
\end{itemize}

\begin{figure*}[ht]
\begin{center}
  \includegraphics[width=.7\textwidth]{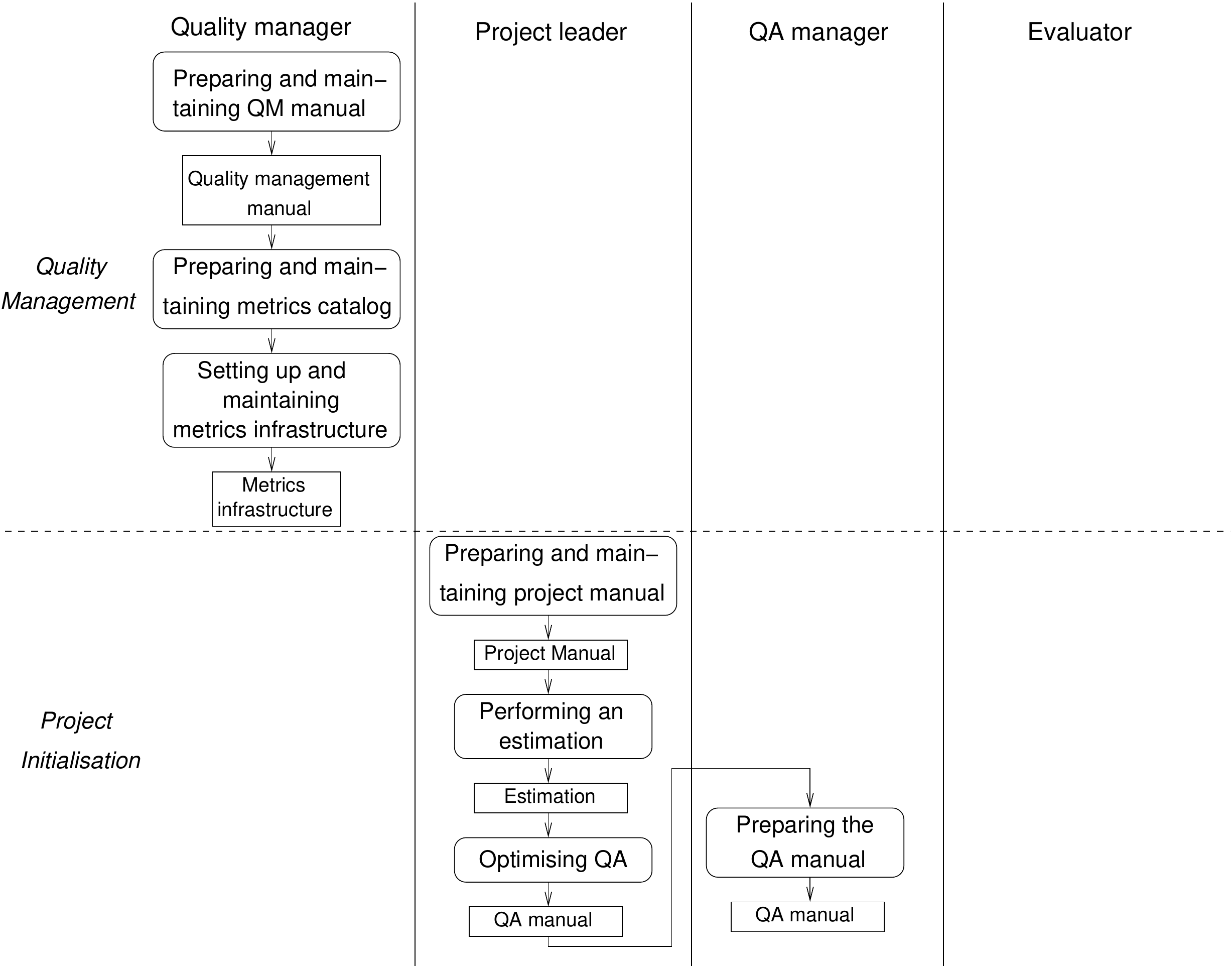}
  \caption{An abstract activity diagram of the model application (first part)}
  \label{fig:activity1}
\end{center}
\end{figure*}

\begin{figure*}[ht]
\begin{center}
  \includegraphics[width=.7\textwidth]{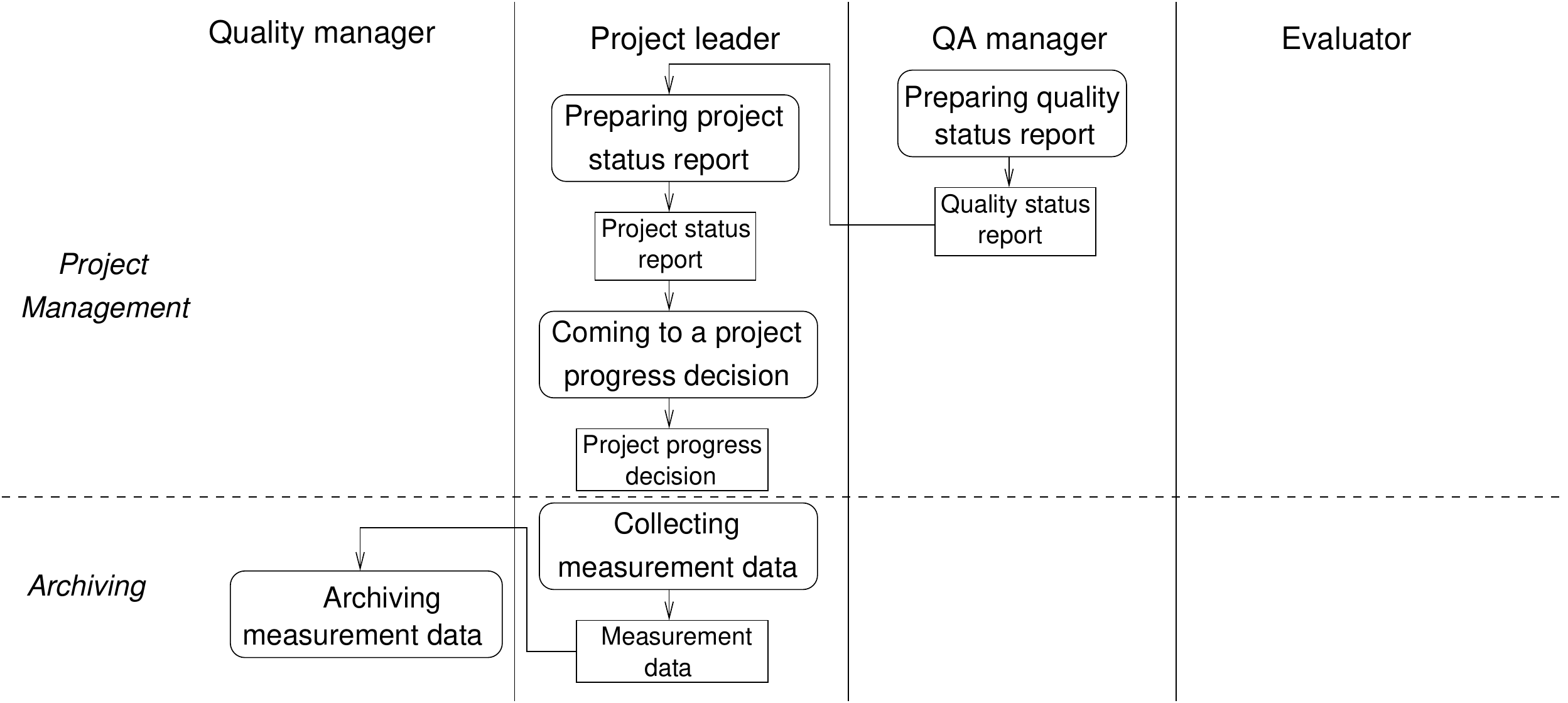}
  \caption{An abstract activity diagram of the model application
           (second part)}
  \label{fig:activity2}
\end{center}
\end{figure*}

\subsection{V-Modell XT Quality Assurance Mechanisms}
\label{sec:vmqa}

Quality assurance is a cornerstone within a process model such as the V-Modell XT. It is the main means to ensure result quality by constructive and analytical means. The QA mechanisms of the V-Modell XT are manifold and cover the following areas:

\begin{enumerate}
 \item \textit{Organisational Quality Management.} Organisational units define general quality standards, guidelines and metrics to be applied in projects. They also archive results in metric catalogues to be used by subsequent projects in order to continuously improve the process on the organisational level. The V-Modell~XT does not define these processes but provides an interface to them. The responsible role is the \emph{Quality Manager}.

	\item \textit{Project Setup.} During project setup, the \emph{Project Manual} and the \emph{QA Manual} are created which define standards, general rules and guidelines that must be applied during the project. The \emph{QA Manual} in particular defines the constructive and analytical methods that are applied for each different type of project result to ensure the desired product quality. Responsible for these tasks are the \emph{Project Leader} and the \emph{QA Manager}.

	\item \textit{Product Evaluation.} Project results are evaluated according to a product specific evaluation specification, following an evaluation procedure. The means applied for evaluating a certain product vary depending on its importance and criticality. \emph{QA Reports} are created in defined intervals describing the state and potential quality problems of the product under construction based on the results of the product evaluations. \emph{Evaluators} are responsible for evaluations and the \emph{QA Manager} for the reports.

	\item \textit{Project Management.} The project management decides about general project progress based on project and \emph{QA Reports}. In case of need, it applies corrective means to increase product quality. Project management reports to the customer and higher levels in the organisation; it is responsible for the overall project result quality as well as for observing agreed upon delivery deadlines and costs. The responsible role is the \emph{Project Leader}.

\end{enumerate}

\subsection{Existing Analytical Process}

The V-Modell~XT already contains an optional process module named ``Measurement and Analysis''. It is very lightweight and only provides the ideas of applying quality analysing metrics within a project. This process module depends only on the core process module of ``Project Management''. Besides, it does not have any other dependencies, and does not extend the general process module ``Quality Assurance''.

In order to keep the compliance level with the standard V-Modell~XT as high as possible, we replace the existing ``Measurement and Analysis'' process module with our own one. However, we reuse the elements defined in the existing process module and complement them with new elements and extended content.

\section{Integration}
\label{sec:integration}

In this section we show the exemplary embedding 
of our model in the process model 
\emph{V-Modell XT}~\cite{v-modell2006}, as described in the previous section.
An integration with other process models
such as the RUP~\cite{kruchten00} can be done accordingly. 
For the V-Modell~XT integration, we define a new process module
``Measurement and Analysis of Analytical QA'' containing roles, products, and activities. We first give
a brief overview of the general analytical QA process and then describe the
contents of the process module in more detail.

\subsection{Overview}

The diagrams shown in Fig.~\ref{fig:activity1} and Fig.~\ref{fig:activity2}
give an overview of the usage of our model as part of the V-Modell XT.
We relate the activities to the different QA processes of the V-Modell~XT
described in Sec.~\ref{sec:vmqa}.

\begin{enumerate}
	\item The \emph{Quality Manager} is responsible for defining and documenting 
cross-project quality standards, 
metrics, and methods, within the \emph{Quality Management Manual}.
Furthermore, he defines and maintains the
\emph{Metrics Catalogue} that must contain all the necessary input factors
summarised in Sec.~\ref{sec:needed}. The \emph{Quality Manager}
must also provide an infrastructure that is able
to store the corresponding metrics.

	\item Within a project, the \emph{Project Leader} is responsible for setting up the
project initially. The \emph{Project Leader} performs
basic estimations for the project including the defect estimate
needed for our model. He can get supporting data from similar projects
using the \emph{Metrics Infrastructure}.
Based on these estimates, he uses our model 
to calculate an optimised quality assurance and
he documents this as part of the \emph{QA Manual}.
Also during project setup, the \emph{QA Manager}
is responsible for implementing the
guidelines from the \emph{Quality Management Manual} within the project.
In collaboration with the \emph{Project Leader} he defines the \emph{QA manual}
and provides input for the QA relevant sections 
of the \emph{Project Manual}. 
The project setup results in a completed \emph{Project Manual} where
project goals and guidelines for supporting project processes, such
as measurement and analysis are defined. 

	\item During the course of the project, \emph{Evaluators} are performing the
product QA tasks. Depending on the specific requirements for
specific work products as expressed in the \emph{QA Manual}, 
they are preparing \emph{Evaluation Specifications} and 
\emph{Evaluation Procedures} for each evaluated product.
The \emph{Project Plan}, under responsibility of project management,
plans the occurrences of the product evaluations and QA measures.
According to these guidelines, the \emph{Evaluators} perform the
actual product evaluations and document the results in
\emph{Evaluation Reports}, which include all measurements 
that can be later used in
our model such as the number and type of the detected defects.
In regular intervals, the \emph{QA Manager} 
compiles \emph{Quality Status Reports}
out of the product \emph{Evaluation Reports}.

	\item The \emph{Project Leader} and project management are continuously
assessing project progress and are responsible for making regular
project progress decisions, which act as ``quality gates''.
One important source of input for
these decisions are the \emph{Quality Status Reports}.
Project management can use the available information to apply
our model to evaluate different scenarios and optimise the
further QA strategy. The results will refine the \emph{Project Plan}
and might lead to an update of the \emph{QA Manual}.
When the project is finished, the project leader collects
the \emph{Measurement Data} relevant for our model and forwards it
to the \emph{Quality Manager} who stores it in the metrics infrastructure.

\end{enumerate}

\subsection{Extensions to the V-Modell~XT}

In the following we show in greater detail the extensions to 
the V-Modell~XT we made to embed our analytical QA model. 
The V-Modell~XT was created as a generic process model
which already puts significant effort on QA and related management
processes. It thus already provides the basic framework for
analytical QA. To fit our model, we have to extend existing elements
of the V-Modell and add a few new ones. Mainly this is only 
more detail on the explicit measurement
and collection of the data for the model input factors.
For our extension, we concentrate on products, activities and roles.

The integration of our analytical quality model affects the
responsibilities of 4 roles that already exist 
in the V-Modell~XT. Their role profile descriptions are
sufficiently abstract and fit our purposes. Thus, only slight
extensions need to be performed:

\begin{itemize}
	\item The \emph{Quality Manager} is responsible for
quality assurance standards across all projects 
and for an efficient and effective quality management
system. 
In particular, he develops a systematic quality management and
creates and maintains the \emph{Quality Management Manual}. Most importantly in
our context, he defines rules and approaches how projects plan and perform
quality assurance techniques. Furthermore, he defines which QA techniques
should be used in general and helps in choosing appropriate techniques
for a specific project.
We add that he is responsible for
setting up and maintaining the \emph{Metrics Infrastructure}.

	\item The \emph{Project Leader}, as the leader of the project's
execution, plans and controls the project's progress.
In particular, he makes the basic estimates for project planning and
decides on future changes based on status reports. The main extension
for this role is that he uses our model for optimising the resource
distribution for quality assurance and also collects the necessary
measurement data for our model.

	\item The \emph{QA Manager} controls the quality in a project and thereby
supervises all quality assurance. He is responsible for the
\emph{Quality Status Reports} and also plans the QA work in collaboration
with others. There is only the small addition that in his 
\emph{Quality Status Report} the necessary measurements for the model must be
contained.

	\item The \emph{Evaluator} -- also called \emph{inspector}
although he not only uses inspections -- creates evaluation specifications
and using those evaluates the artefacts created in the project. Hence,
he uses \DDT s, e.g., reviews and tests, on those artefacts and
reports the results. Also for the \emph{Evaluator} it is necessary that
he documents the necessary measurements for the model factors.

\end{itemize}

Work products are the main V-Modell elements and also the core project
results. Work products have one responsible role.
The following list shows the work products that need to be considered, 
extended or added to apply our model:

\begin{itemize}
	\item The \emph{Quality Management Manual} is a work product that we
add to capture -- among other subjects -- or\-gani\-sa\-tion-wide definitions of 
the metrics that need to be collected for the usage of the model.
Thus \emph{Metrics Definitions} is one subject in this product.
Responsible role for this product is the \emph{Quality Manager}.
The V-Modell~XT mentions such a document but does not officially
introduce it.
	
	\item We also add the \emph{Metrics Catalog}, which exists in the 
V-Modell~XT only as subject of an organisation-wide process model,
for process adaptation and improvement (ORG) projects.
We reuse the subject description and establish it as a full product
under the responsibility of the \emph{Quality Manager}.
We additionally explicitly require the incorporation of the factors 
from Sec.~\ref{sec:needed}.
Thereby, we can reference this product in regular development
projects.

	\item The \emph{Metrics Infrastructure} is the third new work product
under the responsibility of the \emph{Quality Manager}.
It is in essence similar to the existing
\emph{Project Management Infrastructure} but not project specific. In
our context it needs to store the measured data for the relevant
metrics of our model and provide access to it across projects over an
extended period of time.

	\item We extend the existing product \emph{Estimation} with a new 
subject \emph{Estimation of the defect content} that contains data,
which we use later in our model. 
Responsible is the \emph{Project Leader}.

	\item We extend the existing products \emph{Evaluation Report} and
\emph{Quality Status Report} with new subjects that will contain 
the necessary measurement data for the factors of Sec.~\ref{sec:needed}
Responsible are \emph{Evaluator} resp. \emph{QA Manager} roles.

	\item We use the existing work product \emph{Measurement Data} to
capture all data that is collected in the course of the project for
calculating the relevant metrics of our model. Responsible for this
product is the \emph{Project Leader}.

	\item The \emph{Metrics Analysis} is another existing work product
under the responsibility of the \emph{Project Leader}. It contains detailed
analyses of the relevant metrics of our model based on the previously
measured data.
	
\end{itemize}

Each work product has exactly one associated activity.
During execution of such an activity one instance or iteration of the work
product is created or edited. After the completion of the activity,
a product evaluation is performed according the QA guidelines.
The following list explains the relevant activities and sub-activities
to apply our model. Most are related to the given work products above:

\begin{itemize}
	\item We introduce a new activity
\emph{Preparing and maintaining Quality Management Manual} doing what
its name says. A sub-activity describes how the necessary metrics for
applying our model are selected and defined.
	
	\item We introduce the existing sub-activity
\emph{Preparing and maintaining metrics catalogue} of the V-Modell~XT as
full activity creating and maintaining the product \emph{Metrics Catalog}.
	
	\item The new activity \emph{Setting up and maintaining the 
Metrics Infrastructure} will make sure that a data repository 
is available for the measurement data.
	
  \item The main element of our analytical quality assurance model
is introduced into the V-Model~XT as new sub-activity \emph{Optimise QA}.
It belongs to the activity \emph{Preparing the QA Manual}. 
This sub-activity is performed by the \emph{Project Leader} 
with help from the \emph{QA Manager}
based on his estimates and data from similar projects. 
He calibrates the model and optimises
it (w.r.t.\ cost or ROI) so that an optimal resource distribution
is found. This is then documented in the \emph{QA Manual}.
	
	\item The activity \emph{Collecting Measurement Data} describes how the
resp. product is created and edited.
A new sub-activity activity \emph{Archiving Measurement Data} requires
that the measurement data will be stored during the project and at its end 
so that they are available across projects and for new projects.
	
	\item The existing activity \emph{Calculating and Analysing Metrics}
is extended with a new sub-activity to extract the
data that is to be stored for future projects in the \emph{Metrics
Infrastructure}.
	
	\item We add to the activity \emph{Coming to a Project Progress Decision} 
a new sub-activity which uses our economics model as basis 
for the decision. Different scenarios can be analysed and 
an optimal effort (or resource) distribution can be calculated.
		
\end{itemize}

We package all above described new work products and activities as well 
as product/activity extensions in form of subjects and sub-activities
as part of our new process module. We add general descriptions and an
overview of the process module contents.
Thereby we have performed a fully modular 
extension of the V-Model~XT that a \emph{Project Leader} can choose or
not choose to apply during initial V-Modell tailoring for a new project.

\section{Discussion}
\label{sec:discussion}

In summary, we find that our model blends well with the V-Modell~XT.
The necessary changes and additions to use the model fit with the
existing structure and require only additions and slight extensions
of existing  V-Modell~XT elements. They all can be packages nicely as a
process module. An embedding into other
process models with a similar structuring should be possible with
a comparable effort.

One of the challenges of integrating our analytical quality assurance model
into the V-Modell~XT is the scope of the described process. Our model covers
activities on both the organisational level -- spanning multiple projects --
and the project level. 
The main focus of the V-Modell~XT is to describe a process 
for conducting a particular project: once a project is initiated and 
the decision to use the V-Modell~XT is made,
the tailoring activity will result in a project specific process. The 
V-Modell~XT does not specifically cover organisation-wide processes, such as
quality management and continuous process improvement. However, this is not
a limitation, because it can easily be extended -- we have done this for our
integration. This results in a responsibility of the organisation 
to apply the V-Modell~XT process across projects and to provide the necessary
infrastructure.

A consequence of our modular extension of the V-Modell is the sometimes
artificial separation between project specific and organisation specific 
work product (and activity) definitions, such as the \textit{Project Infrastructure} and the 
\textit{Metrics Infrastructure}. Additionally, the existing V-Modell~XT process
support for the \textit{Introduction and Maintenance of Organisation-Specific
Process Models} -- the process module ORG -- covers part of the 
organisation-wide activities without being combinable and integrated with
project specific processes. Thus, we chose to introduce some redundancy
for the sake of modularity and clear understandability.

One of the prerequisites of our approach is the necessity to have 
comparable projects which yield expressive metrics data. The higher
the degree of similarity between the projects and the better the
measured metrics data, the more precise will our model be able to optimise
the QA processes.

\section{Related Work}
\label{sec:related}

General efficiency models of defect-detection techniques such as
the inspection model of Kusumoto et al.\ \cite{kusumoto92} or
the testing model of Morasca and Serra-Capizzano \cite{morasca04}
are aiming at analysing specific techniques and their application.
However, they are typically not usable for planning purposes in
a software project.
Cost models based on reliability models, e.g., Pham \cite{pham00}, aim
to decide when to stop testing. However, they are only applicable
to the system testing phase. 

More economic-oriented models such as iDAVE Boehm et al.\ 
\cite{boehm04} or the model of Slaughter, Harter, and Krishnan 
\cite{slaughter98} are typically more abstract or coarse-grained
than the used model. Moreover, the question when and how to use
those models is not completely clear. Especially, we are not aware
of an integration into an existing process model.

Punter et al.\ \cite{punter04} aim also at a practical application of
product evaluation with specific goals. However, they concentrate
more on the actual evaluation process using the ISO 14598 standard
which we assume given by the V-Modell XT. They also do not explicitly
discuss the aim of using the evaluation results for future optimisations.

Cai et al.\ \cite{cai06} propose a method of optimal and adaptive
testing with cost constraints. They discuss that it is effective to
adapt testing and to explore the interplay between software and control.
However, their model does only consider testing and is not explicitly
integrated in a complete process model.

Ambler uses process patterns in \cite{ambler99} to describe task-specific
self-contained pieces of processes and workflows in a reusable way.
Such patterns can be applied to solve complex tasks when needed.
St\"orrle \cite{stoerrle01} shows how process patterns can be described
in great detail using UML.
The idea of process patterns is further refined by Gnatz et 
al.~\cite{gnatz03} in form of a modular and extensible 
software development process based on collections of independent
process components. These process patterns essentially are the basis
of the extension mechanism of the V-Modell XT.

\section{Conclusions}
\label{sec:conclusions}

Analytical models of quality assurance would be a valuable tool
for project managers and other decision-makers in software projects.
There is a variety of such models available on different levels of
abstraction. However, the adoption in practice is still weak. One
main problem is that the usage of those model is often not clear.
Especially, when and how the model should be used in an existing
process model is typically not specified by the model proposers.

In this paper, we show the exemplary integration of a detailed
model of analytical quality assurance in the process model V-Modell XT.
We are not aware of other models of quality assurance that
have explicitly been integrated into an existing process
model. The benefits of this work are two-fold: (1) organisations
that follow already the V-Modell XT have now simple means to also
incorporate the analytical model into their process. (2) It has
been shown that such an integration can be done relatively simple
and with little effort. Therefore, this should be also possible
with other process models and hence the usage of models of QA can
be increased.

For future work, we consider tool support as the other important
aspect of pushing the use of such models in software organisations.
Hence, we plan to build an easy to use tool implementation that
helps in applying the model. It is also to investigate whether our
claim w.r.t.\ the easy integration into other process models really
holds.

%ACKNOWLEDGMENTS are optional
\section{Acknowledgments}
We are grateful to Ulrike Hammerschall for the discussion on
a first integration. 

%\bibliographystyle{abbrv}
%\bibliography{cost}

\begin{thebibliography}{10}

\bibitem{ambler99}
S.~W. Ambler.
\newblock {\em More Process Patterns: Delivering Large-Scale Systems Using
  Object Technology}.
\newblock Cambridge University Press, 1999.

\bibitem{boehm01}
B.~Boehm and V.~R. Basili.
\newblock {Software Defect Reduction Top 10 List}.
\newblock {\em IEEE Computer}, 34(1):135--137, 2001.

\bibitem{boehm04}
B.~Boehm, L.~Huang, A.~Jain, and R.~Madachy.
\newblock {The ROI of Software Dependability: The iDAVE Model}.
\newblock {\em IEEE Software}, 21(3):54--61, 2004.

\bibitem{v-modell2006}
Bundesrepublik Deutschland.
\newblock {\em V-Modell XT}, 1.2 edition, 2006.
\newblock http://www.v-modell-xt.de/.

\bibitem{cai06}
K.-Y. Cai, Y.-C. Li, W.-Y. Ning, W.~E. Wong, and H.~Hu.
\newblock {Optimal and Adaptive Testing with Cost Constraints}.
\newblock In {\em Proc.\ 2006 International Workshop on Automation of Software
  Test (AST '06)}, pages 71--77. ACM Press, 2006.

\bibitem{chulani99}
S.~Chulani and B.~Boehm.
\newblock {Modeling Software Defect Introduction and Removal: COQUALMO
  (COnstructive QUALity MOdel)}.
\newblock Technical Report USC-CSE-99-510, University of Southern California,
  Center for Software Engineering, 1999.

\bibitem{gnatz03}
M.~Gnatz, F.~Marschall, G.~Popp, A.~Rausch, and W.~Schwerin.
\newblock The living software development process.
\newblock {\em Software Quality Professional}, 5(3), June 2003.

\bibitem{grady92}
R.~B. Grady.
\newblock {\em {Practical Software Metrics for Project Management and Process
  Improvement}}.
\newblock Prentice-Hall, 1992.

\bibitem{jones91}
C.~Jones.
\newblock {\em {Applied Software Measurement: Assuring Productivity and
  Quality}}.
\newblock McGraw-Hill, 1991.

\bibitem{kan02}
S.~H. Kan.
\newblock {\em {Metrics and Models in Software Quality Engineering}}.
\newblock Addison-Wesley, 2nd edition, 2002.

\bibitem{kruchten00}
P.~Kruchten.
\newblock {\em {The Rational Unified Process. An Introduction}}.
\newblock Addison-Wesley, 2nd edition, 2000.

\bibitem{kusumoto92}
S.~Kusumoto, K.~ichi Matasumoto, T.~Kikuno, and K.~Torii.
\newblock {A New Metric for Cost Effectiveness of Software Reviews}.
\newblock {\em IEICE Transactions on Information and Systems},
  E75-D(5):674--680, 1992.

\bibitem{morasca04}
S.~Morasca and S.~Serra-Capizzano.
\newblock {On the Analytical Comparison of Testing Techniques}.
\newblock In {\em Proc.\ 2004 ACM SIGSOFT International Symposium on Software
  Testing and Analysis (ISSTA '04)}, pages 154--164. ACM Press, 2004.

\bibitem{myers79}
G.~J. Myers.
\newblock {\em {The Art of Software Testing}}.
\newblock John Wiley \& Sons, 1979.

\bibitem{pham00}
H.~Pham.
\newblock {\em {Software Reliability}}.
\newblock Springer, 2000.

\bibitem{punter04}
T.~Punter, R.~Kusters, J.~Trienekens, T.~Bemelmans, and A.~Brombacher.
\newblock {The W-Process for Software Product Evaluation: A Method for
  Goal-Oriented Implementation of the ISO 14598 Standard}.
\newblock {\em Software Quality Journal}, 12(2):137--158, 2004.

\bibitem{rai98}
A.~Rai, H.~Song, and M.~Troutt.
\newblock {Software Quality Assurance: An Analytical Survey and Research
  Prioritization}.
\newblock {\em Journal of Systems and Software}, 40:67--83, 1998.

\bibitem{Rausch2006}
A.~Rausch and M.~Broy.
\newblock {\em {Das V-Modell XT. Grundlagen, Erfahrungen und Werkzeuge}}.
\newblock dpunkt.verlag, 2006.
\newblock In German.

\bibitem{nist02}
RTI.
\newblock {The Economic Impacts of Inadequate Infrastructure for Software
  Testing}.
\newblock Planning Report 02--3, National Institute of Standards \& Technology,
  2002.

\bibitem{slaughter98}
S.~A. Slaughter, D.~E. Harter, and M.~S. Krishnan.
\newblock {Evaluating the Cost of Software Quality}.
\newblock {\em Communications of the ACM}, 41(8):67--73, 1998.

\bibitem{song06}
Q.~Song, M.~Sheperd, M.~Cartwright, and C.~Mair.
\newblock {Software Defect Association Mining and Defect Correction Effort
  Prediction}.
\newblock {\em IEEE Transactions on Software Engineering}, 32(2):69--82, 2006.

\bibitem{stoerrle01}
H.~St\"orrle.
\newblock {Describing Process Patterns with UML}.
\newblock In {\em Proc.\ 8th European Workshop on Software Process Technology
  (EWSPT '01)}, pages 173--182. Springer, 2001.

\bibitem{wagner:isese06}
S.~Wagner.
\newblock {A Literature Survey of the Quality Economics of Defect-Detection
  Techniques}.
\newblock In {\em Proc.\ 5th ACM-IEEE International Symposium on Empirical
  Software Engineering (ISESE '06)}. ACM Press, 2006.

\bibitem{wagner:issta06}
S.~Wagner.
\newblock {A Model and Sensitivity Analysis of the Quality Economics of
  Defect-Detection Techniques}.
\newblock In {\em Proc.\ International Symposium on Software Testing and
  Analysis (ISSTA '06)}, pages 73--83. ACM Press, 2006.

\bibitem{wagner:diss06}
S.~Wagner.
\newblock {\em {Cost-Optimisation of Analytical Software Quality Assurance}}.
\newblock {PhD Dissertation}, Technische Universit\"at M\"unchen, 2006.
\newblock To appear.

\end{thebibliography}
% ACM needs 'a single self-contained file'!

\balancecolumns

\end{document}